\documentclass[floatfix]{revtex4}
\usepackage{amsmath}
\usepackage{color}
\usepackage{amssymb}

\usepackage{graphicx}

\begin{document}

\title{Theory of type-II superconductivity in ferromagnetic metals with triplet pairing}

\author{V.P.Mineev$^{1,2}$}
\affiliation{$^1$Universite Grenoble Alpes, CEA, INAC, PHELIQS, GT, F-38000 Grenoble, France\\
$^2$Landau Institute for Theoretical Physics, 142432 Chernogolovka, Russia}

\begin{abstract}
The superconducting state in uranium compounds UGe$_2$, URhGe and UCoGe is formed at temperatures far below the Curie temperature
pointing on nonconventional nature of superconductivity in these materials - namely the superconductivity with triplet pairing.
The emergence of superconductivity is accompanied by the slight magnetization expulsion typical for the type-II superconductors. 
Following classic Abrikosov paper I  develop the theory of type-II superconductivity in application to two-band ferromagnetic metal with equal spin triplet pairing.
\smallskip

\end{abstract}

\date{\today}
\maketitle
\section{Introduction}
The investigations of interplay between superconductivity and magnetism have long story. Usually ferromagnetic ordering suppresses the superconducting state because the exchange field exceeds the paramagnetic limit field and aligns the electron spins directed oppositely in Cooper pairs. Nevertheless, singlet superconductivity can coexist with ferromagnetism when the critical temperature of the transition to the superconducting state is greater than the Curie temperature, as is the case with ternary compounds investigated in the 1980s (for review see \cite{Maple1995}).  The coexistence occurs in a form crypto-ferromagnetic superconducting state characterized by appearance a  periodic magnetic structure with period larger than the interatomic distance, but smaller than the superconducting coherence length, which weakens the depairing effect of the exchange field.

The superconductivity in the more recently discovered uranium compounds UGe$_2$, URhGe and UCoGe   \cite{Saxena01,Aoki01,Huy07} exhibits quite different properties
(see  the experimental \cite{Aoki2014} and theoretical reviews \cite{Mineev2016} and references therein). 
 Here the superconducting states exist at temperatures far below the Curie temperature  Fig.1 
 and in the magnetic fields strongly exceeding the paramagnetic limit indicating that we deal with the triplet pairing.  
 The general form of superconducting order parameters in these orthorhombic compounds is found in the paper \cite{Mineev2002}.
 Similar to the superfluid $^3$He the pairing interaction  is caused by the magnetic fluctuations. The theory based on this mechanism and on the symmetry considerations allows explain many specific properties of these materials \cite{Mineev2016}.

 Quite recently there was proposed the phenomenological description of the phase diagram of UCoGe \cite{Raghu2016,Mineev2017} where the ferromagnetism is suppressed by pressure whereas the superconductivity arising at small pressures inside of the ferromagnetic state continues to exist at high pressures in the paramagnetic state Fig.1c. The theory was developed as if it would be in the neutral superfluid.  This approach is justified by the smallness of the internal  magnetic field  interacting with the electron charges  
  that   slightly changes  the critical temperature of transition to the superconducting state. The effects  caused by the screening supercurrents has been taken into account  only qualitatively \cite{Mineev2017}. This has  allowed to explain the significant difference between the transition from the ferromagnetic to the ferromagnetic superconducting state and the transition from the superconducting to the ferromagnetic superconducting state. However, the developed theory was not completely consistent. 
  
  All the aforementioned superconductors are related to the  type-II superconducting materials \cite{Dao2011}. The 
  internal magnetic fields in all of them 
  exceed the corresponding lower critical fields $H_{c1}$  \cite{Aoki2014, Deguchi2010, Paulsen2012, Hykel2014} . Hence, at temperature decrease  the phase transition from the ferromagnetic to the ferromagnetic superconducting state  occurs to the  mixed state characterized by the emergence of Abrikosov vortices. Accordingly, the proper theory of
  this phase transition must be formulated in frame  Ginzburg-Landau-Abrikosov theory of type-II superconductivity \cite{Abrikosov1957}. In the application to ferromagnetic
  conventional superconductors with singlet pairing such approach has been   developed first in the papers \cite{Blount1979, Kuper1980}.
  The corresponding theory for the nonmagnetic superconductors with equal spin triplet pairing in absence of spin-orbital coupling has been presented in the paper \cite{Agterberg2009}.
  
  Here, I develop the Abrikosov theory of type-II superconductivity  for equal spin pairing triplet superconducting state in two band ferromagnetic metal.  First, I describe the phase transition from ferromagnetic to superconducting ferromagnetic state that occurs in all three uranium compounds (see Fig.1). Then I consider the solution for isolated vortex in  such type superconductors and
the transition from the Meissner to the mixed superconducting state which is realized in UCoGe. 
  In my derivation I use the pedagogic  presentation of  classic Abrikosov theory  performed by N.B.Kopnin \cite{Kopnin2002}.

  \section{Model}

The triplet-pairing superconducting state order parameter in two-band (spin-up, spin-down) ferromagnet is given by the complex spin-vector \cite{Mineev2016,Book} 
\begin{eqnarray}
{\bf d}({\bf k},{\bf r})=~~~~~~~~~~~~~~~~~~~~~~~~~~~\\
\frac{1}{2}
\left[-\Delta^{\uparrow}¥({\bf k},{\bf r})(\hat{x}+i\hat{y})+
\Delta^{\downarrow}¥({\bf k},{\bf r})(\hat{x}-i\hat{y})\right]+\Delta^{0}¥({\bf k},{\bf r})\hat z,\nonumber
\label{e14}
\end{eqnarray}
where $\Delta_{\uparrow}¥({\bf k},{\bf r})$,  $\Delta_{\downarrow}¥({\bf r},{\bf k},{\bf r})$, $ \Delta_{0}¥({\bf k},{\bf r}) $ are the amplitudes of spin-up, spin-down and zero-spin  of superconducting order parameter depending on the Cooper pair centre of gravity coordinate ${\bf r}$ and the common direction of momentum ${\bf k }$ of pairing electrons.  In the orthorhombic ferromagnets with easy axis along $\hat z$ direction
there are only two superconducting states  A and B  with different critical temperature  \cite{Mineev2002}.  
We will work with equal spin pairing B-state with the order parameter
\begin{eqnarray}
&\Delta_B^\uparrow({\bf k},{\bf r})=\hat k_z\eta_1({\bf r}),\nonumber\\
&\Delta_B^\downarrow({\bf k},{\bf r})=\hat k_z\eta_2({\bf r}).
\label{B}
\end{eqnarray}

The Ginzburg-Landau free energy functional is
\begin{eqnarray}
&F=\int dV\left\{\alpha M^2+\beta M^4 +D_{ij}\nabla_iM\nabla_jM\right.\nonumber\\
&+\left.\alpha_1(|\eta_1|^2+|\eta_2|^2)+\gamma_1({\bf B}\hat z)(|\eta_1|^2-|\eta_2|^2)+
\gamma_2(\eta_1\eta_2^\star+\eta_1^\star\eta_2)+
\beta_1(|\eta_1|^4+|\eta_2|^4)+
\beta_2|\eta_1|^2|\eta_2|^2\right.\nonumber\\
&\left.+K_{1ij}(D_i\eta_1)^\star D_j\eta_1+K_{2ij}(D_i\eta_2)^\star D_j\eta_2+\frac{\bf B^2}{8\pi}-{\bf B}{\bf M}\right\},
\label{FEF}
\end{eqnarray}
where $M$ is the density of magnetic moment component along the easy axis, ${\bf B}=curl{\bf A}$ is the magnetic induction,
\begin{equation}
\alpha=\alpha_0(T-T^\prime_c),~~~~~\alpha_1=\alpha_{10}(T-T_{sc0}),
\end{equation}
$T_c^\prime( P )$ is the pressure dependent "Curie temperature" ( see \cite{Blount1979}) and $T_{sc0}( P )$ is the formal  critical temperature of superconducting transition in the single band (say just spin-up) case.
${\bf D}=-i\hbar\nabla-\frac{2e}{c}{\bf A}$ is the long derivative. 
 In  a single domain ferromagnet in the absence of external field $H=0$ or at the external field directed along the axis of spontaneous magnetization $\hat z$
the order parameter components are the $z$-coordinate independent and the long derivatives are
\begin{equation}
D_x=-i\hbar\frac{\partial}{\partial x},
~~
D_y=-i\hbar\frac{\partial}{\partial y}-\frac{2e}{c}A_y.
\end{equation}
For the superconducting state (\ref{B}) the gradient terms have the following form $$K_{1xx}(D_x\eta_1)^\star D_x\eta_1+K_{1yy}(D_y\eta_1)^\star D_y\eta_1+(1~\to~2).$$
The upper critical field problem in two band superconductor with different
  stiffness constants $K_{1xx}$ and  $K_{1yy}$ 
   can be solved only numerically  or by means of variation approach used in the paper by Zhitomirsky and  Dao \cite{Dao2004}.
  With purpose to develop the analytic treatment we neglect  the orthorhombicity puting $K_{1xx}=K_{1yy}=K_1$, $K_{2xx}=K_{2yy}=K_2$ and
  also $D_{xx}=D_{yy}=D$.
  
An analytic solution can be found also for the equal spin pairing A-state   
 \begin{eqnarray}
&\Delta^\uparrow({\bf k},{\bf r})=\hat k_x\eta_1({\bf r})
,\nonumber\\
&\Delta^\downarrow({\bf k},{\bf r})=\hat k_x\eta_2({\bf r}).
\label{A}
\end{eqnarray}
discussed in the papers \cite{Raghu2016,Mineev2017}. Then, however, due to   the gradient mixing terms like
$(D_x\eta_{1x})^\star D_y\eta_{1y}$ the order parameter (\ref{A}) acquires  (see\cite{Mineev2016}) more general form 
\begin{eqnarray}
&\Delta^\uparrow({\bf k},{\bf r})=\hat k_x\eta_{1x}({\bf r})+i\hat k_y\eta_{1y}({\bf r})
,
\label {Aup}\\
&\Delta^\downarrow({\bf k},{\bf r})=\hat k_x\eta_{2x}({\bf r})+i\hat k_y\eta_{2y}({\bf r}).
\label{Adown}
\end{eqnarray}
Thus, instead two GL equations for the superconducting order parameters one has to solve four of them. The linear equations for $\eta_{1x}, \eta_{1y},\eta_{2x},\eta_{2y}$ can be solved making use  the generalization on two band case  the problem of the upper critical field   in uniaxial superconductor with two-component order parameter under magnetic field directed along four-fold axis (see \cite{Book}). This, however, leads to very cumbersome equations and we prefer to work with the state given by Eq.(\ref{B}) and the free energy functional
\begin{eqnarray}
&F=\int dV\left\{\alpha M^2+\beta M^4 +D(\nabla_xM)^2+D(\nabla_yM)^2+\right.\nonumber\\
&+\left.\alpha_1(|\eta_1|^2+|\eta_2|^2)+\gamma_1({\bf B}\hat z)(|\eta_1|^2-|\eta_2|^2)+
\gamma_2(\eta_1\eta_2^\star+\eta_1^\star\eta_2)+
\beta_1(|\eta_1|^4+|\eta_2|^4)+
\beta_2|\eta_1|^2|\eta_2|^2\right.\nonumber\\
&\left.+K_{1}[(D_x\eta_1)^\star D_x\eta_1+(D_y\eta_1)^\star D_y\eta_1]+K_{2}[(D_x\eta_2)^\star D_x\eta_2+(D_y\eta_2)^\star D_y\eta_2]
+\frac{\bf B^2}{8\pi}-{\bf B}{\bf M}\right\}.
\label{FE}
\end{eqnarray}

\section{Transition from  ferromagnetic to superconducting ferromagnetic state}

In URhGe and UCoGe below phase transition in ferromagnetic state
the magnetic moment acquires the finite value, the magnetic induction is $B=4\pi M$ and a superconducting ordering is absent
\begin{equation}
M^2=(M_0(T))^2=-\frac{\alpha_0(T-T_{c}( P ))}{2\beta},~~~\eta_1=\eta_2=0,
\label{M}
\end{equation}
where the Curie temperature is
\begin{equation}
T_c=T_c^\prime+\frac{2\pi}{\alpha_0}.
\end{equation}
In presence of an external field $H=B-4\pi M$ parallel to spontaneous magnetization the magnetic moment is determined by the equation
\begin{equation}
2\alpha M_0+4\beta M_0^3=H.
\label{MH}
\end{equation}
At arbitrary temperatures below the Curie temperature, one can work with the GL formula for $M_0$  only qualitatively. Instead,  it is possible to use the known experimental values of  magnetization $M_0(H,T)$
The same is true for
UGe$_2$ where the superconductivity arises below the first order phase transition to ferromagnetic state Fig.1. 

At the subsequent phase transition the superconducting order parameter amplitudes $\eta_1, \eta_2$ appear. They are determined by the Ginzburg-Landau equations obtained by variation of Eq.(\ref{FE}) in respect to $\eta_1, \eta_2$
\begin{eqnarray}
(\alpha_1+\gamma_1B)\eta_1-K_1\left [ \frac{\partial^2}{\partial x^2}+\left (\frac{\partial}{\partial y}-\frac{2ieB}{\hbar c}x\right )^2 \right ]\eta_1+\gamma_2\eta_2+2\beta_1|\eta_1|^2\eta_1
+\beta_2\eta_1|\eta_2|^2=0,
\label{GL1}
\\
\gamma_2\eta_1+(\alpha_1-\gamma_1B)\eta_2-K_2\left [ \frac{\partial^2}{\partial x^2}+\left (\frac{\partial}{\partial y}-\frac{2ieB}{\hbar c}x\right )^2 \right ]\eta_2+2\beta_1|\eta_2|^2\eta_2+
\beta_2|\eta_1|^2\eta_2=0.
\label{GL2}
\end{eqnarray}

\subsection{Upper critical field}

The   transition to the superconducting state  occurs at $B_{c2}(T)$ which is the eigen value
of the corresponding linear equations
\begin{eqnarray}
(\alpha_{1}+\gamma_1B_{c2})\eta_{10}-K_1\left [ \frac{\partial^2}{\partial x^2}+\left (\frac{\partial^2}{\partial y}-\frac{2ieB_{c2}}{\hbar c}x\right )^2 \right ]\eta_{10}+\gamma_2\eta_{20}=0,
\label{linear1}\\
\gamma_2\eta_{10}+(\alpha_{1}-\gamma_1B_{c2})\eta_{20}-K_2\left [ \frac{\partial^2}{\partial x^2}+\left (\frac{\partial^2}{\partial y}-\frac{2ieB_{c2}}{\hbar c}x\right )^2 \right ]\eta_{20}=0.
\label{linear2}
\end{eqnarray}
The solution of this system for the lowest eigen value
 is
 \begin{equation}
\eta_{i0}= C_{i}\exp\left[-\frac{\pi B_{c2}}{\Phi_0}\left (x-\frac{q\Phi_0}{2\pi B_{c2}}\right)^2 \right],~~~~~~                    i=1,2,
\label{L1}
\end{equation}
 where $\Phi_0=\frac{\pi\hbar c}{e}$ is the magnetic flux quantum.
Substitution of   solutions  back to equations yields the system of linear equations for coefficients $C_1,C_2$. The equality of  the determinant of this system to zero yields the equation for the $B_{c2}(T)$ 
\begin{equation}
\left(\frac{2\pi B_{c2}}{\Phi_0}\right )^2+ \left(\frac{\alpha_1+\gamma_1B_{c2}}{K_2} -\frac{\alpha_1-\gamma_1B_{c2}}{K_1} \right )\frac{2\pi B_{c2}}{\Phi_0} +\frac{\alpha_1^2-(\gamma_1B_{c2})^2-\gamma_2^2}{K_1K_2}=0.
\label{CI}
\end{equation}
It contains the terms $\alpha_1\pm\gamma_1B_{c2}=\alpha_{10}(T-T_{sc0}\pm\gamma_1B_{c2}/\alpha_{10})$ corresponding to the shifts of critical temperature in spin-up and spin-down bands. In a magnetic  (nonunitary) superconducting state the shift of $T_{sc0}$ is much smaller than the temperature $T_{sc0}$ (see\cite{Book}):
\begin{equation}
\frac{\gamma_1B_{c2}}{\alpha_{10}}\approx\frac{\mu_BB_{c2}}{\varepsilon_F}T_{sc0},
\label{gamma1}
\end{equation}
where $\mu_B$ is the Bohr magneton and $\varepsilon_F$ is the Fermi energy. In neglect  these terms
\begin{equation}
B_{c2}(T)=\frac{\Phi_0}{2\pi}\left\{-\frac{\alpha_1}{2K_2}-\frac{\alpha_1}{2K_1}+\left[
\left (\frac{\alpha_1}{2K_2}-\frac{\alpha_1}{2K_1}\right )^2+\frac{\gamma_2^2}{K_1K_2}   \right ]^{1/2}\right\}.
\label{Bc2}
\end{equation}
In the absence of external field the ferromagnet volume is filled by the domains with opposite magnetization orientation and the equation 
\begin{equation}
B_{c2}(T_{sc})=4\pi M_0(T_{sc}).
\end{equation}
determines the critical temperature $T_{sc}$ of transition to the superconducting state.
When the external field increases the parallel to the field domains are expanded,  the antiparallel domains are shrunk and the critical temperature does not change till 
$H=4\pi M_0$  \cite{Hardy2005}.
When the external field exceeds $4\pi M_0$ the multi-domain ferromagnetic  structure is suppressed.
We will develop theory  for phase transition to superconducting state in single ferromagnetic domain with magnetization parallel to the external field
where the upper critical field at temperatures below $T_{sc}$ is determined by equation
\begin{equation}
H_{c2}(T)= B_{c2}(T)-4\pi M_0(T),
\end{equation}
that near the critical temperature is
\begin{equation}
H_{c2}(T)=\frac{\partial (B_{c2}(T)-4\pi M_0(T))}{\partial T} |_{T=T_{sc}}(T-T_{sc}).
\end{equation}
One must remember, however, that the actual upper critical field in multi-domain specimen at given temperature $T<T_{sc}$ is shifted up 
on $4\pi M$ in respect to this value (see Fig.2).

I will not write the explicit formula for $T_{sc}$ and for $\frac{\partial B_{c2}(T)}{\partial T}\left |_{T=T_{sc}}\right.$. They are quite cumbersome even in negligence of temperature and field dependence of magnetization $M_0$. A reader can easily obtain them.

\subsection{Vortex lattice}

The solution  (\ref{L1}) is centered at $x=\hbar cq/2eB_{c2}=q\Phi_0/2\pi B_{c2}$.
The full solution is a linear combination of these solutions for different $q$. One can construct a periodic solution of the form
\begin{equation}
\eta_{i0}= \sum_n C_{i,n}\exp\left[iqny-\frac{\pi B_{c2}}{\Phi_0}\left (x-\frac{qn\Phi_0}{2\pi B_{c2}}\right)^2 \right],~~~~~~                    i=1,2.
\label{L}
\end{equation}

It is periodic in $y$ with period $Y_0=2\pi/q$. It would be periodic in  $x$ as well if the coefficients satisfy the periodicity condition $C_{i,n+p}=C_{i,n}$,
where $p$ is an integer.
Then,
\begin{equation}
\eta_{i0}\left(x+\frac{p\hbar c q}{2eB_{c2}},y\right)=e^{ipqy}\eta_{i0}(x,y)
\end{equation}
The simplest case is realized when all the coefficients $C_{i,n}=C_i$ are $n$-independent. The array forms a rectangular lattice.

The modulus of these distributions are double periodic with periods 
$$
X_0=\frac{\hbar cq}{2eB_{c2}},~~~~Y_0=\frac{2\pi}{q}.
$$
The unit cell area of rectangular lattice is
\begin{equation}
X_0Y_0=\frac{\Phi_0}{B_{c2}}=2\pi\xi^2,
\end{equation}
which corresponds to exactly one flux quantum per unit cell. If $q$ is chosen in such a way that $X_0=Y_0$, we obtain a square lattice.

\subsection{Magnetization decrease below transition to the superconducting ferromagnetic state}

 At magnetic field $H$ slightly below  $H_{c2}(T)$ 
 there is the screening of magnetization by superconducting currents. 
 This case the superconducting order parameter amplitudes and 
the ferromagnetic  moment acquire the small correction 
\begin{equation} 
\eta_1=\eta_{10}+\tilde\eta_1,~~~~\eta_2=\eta_{20}+\tilde\eta_2,~~~~~{\bf M}=M_0\hat z+m({\bf r})\hat z.
\end{equation}
The same is true for the vector-potential which is
\begin{eqnarray}
{\bf A}={\bf A}_0+{\bf A}_1,~~~~{\bf A}_0=(0,B_{c2}x,0),~~~~ {\bf A}_1=(0,(H+4\pi M_0-B_{c2})x,0)+\delta{\bf A}({\bf r}).
\label{vecpot}
\end{eqnarray}
The corresponding magnetic induction is
\begin{equation}
{\bf  B}=curl{\bf A}==(H+4\pi M_0)\hat z+\delta B({\bf r})\hat z.
\label{Ind}
\end{equation}
It is important to note that in
the ferromagnetic superconducting mixed state the  specimen magnetization is not equal to $M=M_0+m({\bf r})$ but
\begin{equation}
{\cal M}=\frac{\langle( B-H)\rangle}{4\pi}=M_0+\frac{\langle\delta B({\bf r})\rangle}{4\pi}.
\label{moment}
\end{equation}
where  $\langle(\dots)\rangle=S^{-1}\int dxdy(\dots)$ is the space average
over the surface perpendicular to spontaneous magnetization.

By variation of the functional Eq.(\ref{FE}) in respect to the vector potential we obtain the Maxwell equation
\begin{equation}
\frac{c}{4\pi}curl[\delta{\bf B}-4\pi{\bf m}+4\pi\hat z\gamma_1(|\eta_{10}|^2-|\eta_{20}|^2)]={\bf j}=-\frac{2ie}{\hbar c} K_1\left [\eta_{10}^\star(\nabla-\frac{2ie}{\hbar c}{\bf A}_0) \eta_{10}- 
\eta_{10}(\nabla+\frac{2ie}{\hbar c}{\bf A}_0) \eta_{10}^\star  \right ]+(1\to 2)
\end{equation} 
or
\begin{eqnarray}
&\frac{c}{4\pi}\frac{\partial[ \delta B-4\pi m+4\pi\gamma_1(|\eta_{10}|^2-|\eta_{20}|^2)]}{\partial y}=j_x=-\frac{2e}{\hbar}\left\{K_1\left(i\eta_{10}^\star\frac{\partial \eta_{10}}{\partial x}-i\eta_{10}\frac{\partial \eta_{10}^\star}{\partial x}  \right ) +(1\to 2)  \right \},\\
&-\frac{c}{4\pi}\frac{\partial[\delta B-4\pi m+4\pi\gamma_1(|\eta_{10}|^2-|\eta_{20}|^2)]}{\partial x}=j_y=-\frac{2e}{\hbar}\left \{ K_1\left(\eta_{10}^\star(i\frac{\partial}{\partial x}+\frac{2\pi B_{c2}}{\Phi_0}) \eta_{10}-\eta_{10}(i\frac{\partial }{\partial x}-\frac{2\pi B_{c2}}{\Phi_0})\eta_{10}^\star  \right ) +(1\to 2)  \right \}.
\end{eqnarray}
With help of relation
$$
\frac{\partial \eta_{i0}}{\partial x}=\left(-i\frac{\partial}{\partial y}-\frac{2\pi B_{c2}}{\Phi_0}\right )\eta_{i0}
$$
one can rewrite the Maxwell equations as
\begin{eqnarray}
\frac{\partial [\delta B-4\pi m+4\pi\gamma_1(|\eta_{10}|^2-|\eta_{20}|^2)]}{\partial y}=-4\pi\frac{2\pi}{\Phi_0}\left(K_1\frac{\partial |\eta_{10}|^2}{\partial y}+
K_2\frac{\partial |\eta_{20}|^2}{\partial y}\right ),\\
\frac{\partial[\delta B-4\pi m+4\pi\gamma_1(|\eta_{10}|^2-|\eta_{20}|^2)]}{\partial x}=-4\pi\frac{2\pi}{\Phi_0}\left (K_1\frac{\partial |\eta_{10}|^2}{\partial x}+
K_2\frac{\partial |\eta_{20}|^2}{\partial x}\right ).
\end{eqnarray}
Hence,
\begin{equation}
\delta B=-4\pi\frac{2\pi}{\Phi_0}\left(K_1|\eta_{10}|^2+K_2|\eta_{20}|^2\right )+4\pi m-4\pi\gamma_1(|\eta_{10}|^2-|\eta_{20}|^2).
\label{deltaB}
\end{equation}

Now, let us find  $m({\bf r}) $.
Below $T_{sc}$ the magnetization  is determined from the equation
\begin{equation}
2\alpha M+4\beta M^3
 -2D\Delta M-B=0,
\end{equation}
obtained by the variation of the  functional Eq.(\ref{FE}) in  respect to $M$. Here $\Delta$ is the 2D Laplacean. 
Hence, the correction to magnetization $m=M-M_0$ is determined by the equation
\begin{equation}
(2\alpha +12\beta M_0^2-2D\Delta)m
=\delta B
\end{equation}
which, taking into account Eq.(\ref{deltaB}), can be rewritten as
\begin{equation}
(2\tilde\alpha +12\beta M^2-2D\Delta)m
=-4\pi\frac{2\pi}{\Phi_0}\left(K_1|\eta_{10}|^2+
K_2|\eta_{20}|^2\right )-4\pi\gamma_1(|\eta_{10}|^2-|\eta_{20}|^2),
\end{equation}
where $\tilde\alpha=\alpha_0(T-T_c)=\alpha_0(T-T_c^\prime-2\pi/\alpha_0)$.
The magnetic coherence length is much shorter than the 
size of vortex lattice cell
$$
\xi_m=\frac{\sqrt{D}}{\sqrt{\tilde\alpha +6\beta M^2}}<<\xi.
$$
Hence, 
\begin{equation}
m({\bf r})=-\frac{\frac{(2\pi)^2}{\Phi_0}\left(K_1|\eta_{10}|^2+
K_2|\eta_{20}|^2\right )+2\pi\gamma_1(|\eta_{10}|^2-|\eta_{20}|^2)
}{\tilde\alpha +6\beta M_0^2}.
\label{m2}
\end{equation}

According to Eq.(\ref{moment}) the magnetization decrease below transition to the superconducting state is
\begin{equation}
{\cal M}-M_0=\frac{\langle\delta B({\bf r})\rangle}{4\pi}=-\left\langle\frac{2\pi}{\Phi_0}\left(K_1|\eta_{10}({\bf r})|^2+K_2|\eta_{20}({\bf r})|^2\right )-m({\bf r}) +\gamma_1(|\eta_{10}|^2-|\eta_{20}|^2)\right\rangle.
\label{moment1}
\end{equation}
In absence of an external field this  space average $\propto(T_{sc}-T)$. It can be calculated 
substituting the functions $\eta_{10},\eta_{20}$ in the GL functional and then finding its stationary solutions in respect of constant $C_1$ and $C_2$
 at $B=4\pi M_0$. For phase transition in an external field one can   express this average through the difference $H_{c2}-H$
like it was done in the classic Abrikosov paper \cite{Abrikosov1957}.

To find the average we are searching for
let us write the GL equations (\ref{GL1}), (\ref{GL2}) in the matrix form
\begin{equation}
\left (\begin{array}{c}
\alpha _1+\gamma_1B -K_1(\nabla-\frac{2ie}{\hbar c}{\bf A})^2 ;~~~\gamma_2 \\
 \gamma_2;~~~\alpha _1-\gamma_1B -K_2(\nabla-\frac{2ie}{\hbar c}{\bf A})^2 \end{array}\right )\left (\begin{array}{c}\eta_1\\
 \eta_2\end{array}\right ) +\left (\begin{array}{c}(2\beta_1|\eta_1|^2+\beta_2|\eta_2|^2)\eta_1\\
 (2\beta_1|\eta_2|^2+\beta_2|\eta_1|^2)\eta_2 \end{array}\right )
 =\left (\begin{array}{c}0\\
0\end{array}\right ).
\end{equation}
Using the corresponding linear equations ({\ref{linear1}),(\ref{linear2}) one can obtain the equations for the small corrections
\begin{eqnarray}
&\left (\begin{array}{c}
\alpha _1+\gamma_1B_{c2} -K_1(\nabla-\frac{2ie}{\hbar c}{\bf A}_0)^2 ;~~~\gamma_2 \\
 \gamma_2;~~~\alpha _1-\gamma_1B_{c2} -K_2(\nabla-\frac{2ie}{\hbar c}{\bf A}_0)^2 \end{array}\right )\left (\begin{array}{c}\tilde\eta_1\\
 \tilde\eta_2\end{array}\right )-\frac{2ie}{\hbar c}
 \left (\begin{array}{c}K_1{\bf A}_1(\nabla-\frac{2ie}{\hbar c}{\bf A}_0) \eta_{10}+K_1(\nabla-\frac{2ie}{\hbar c}{\bf A}_0){\bf A}_1 \eta_{10}\\
 K_2{\bf A}_1(\nabla-\frac{2ie}{\hbar c}{\bf A}_0) \eta_{20}+K_2(\nabla-\frac{2ie}{\hbar c}{\bf A}_0){\bf A}_1 \eta_{20}\end{array}\right )\nonumber\\
&+\gamma_1(H-H_{c2}+\delta B)
  \left (\begin{array}{c}\eta_{10}\\
 - \eta_{20}\end{array}\right )  +\left (\begin{array}{c}(2\beta_1|\eta_{10}|^2+\beta_2|\eta_{20}|^2)\eta_{10}\\
 (2\beta_{10}|\eta_{20}|^2+\beta_2|\eta_1|^2)\eta_{20} \end{array}\right ) =\left (\begin{array}{c}0\\
0\end{array}\right )
\end{eqnarray}
Let us multiply this column from the left on the line $(\eta_{10}^\star,~\eta_{20}^\star)$ and integrate the obtained product over the surface perpendicular to spontaneous magnetization $\langle(É)\rangle=S^{-1}\int dxdy(É)$. Then after integrating by parts  we find that the integral from the first term  in the product is equal to zero and the other terms are collected into the following expression
\begin{equation}
\left \langle-\frac{1}{c}({\bf j}{\bf A}_1)+\gamma_1(H-H_{c2}+\delta B)(|\eta_{10}|^2-|\eta_{20}|^2)+2\beta_1(|\eta_{10}|^4+|\eta_{20}|^4)+2\beta_2|\eta_{10}|^2|\eta_{20}|^2
\right \rangle=0.
\end{equation}
The current density is ${\bf j}=\frac{c}{4\pi}curl(\delta{\bf B}-4\pi{\bf m}+4\pi\hat z\gamma_1(|\eta_{10}|^2-|\eta_{20}|^2))$. Integrating the first term by parts we obtain
\begin{eqnarray}
\left \langle-\frac{1}{4\pi}(\delta B-4\pi m
)(H-H_{c2}+\delta B)
+2\beta_1(|\eta_{10}|^4+|\eta_{20}|^4)+2\beta_2|\eta_{10}|^2|\eta_{20}|^2     \right \rangle=0,
\end{eqnarray}
and using (\ref{deltaB})
\begin{eqnarray}
&\left \langle\frac{2\pi}{\Phi_0}\left(K_1|\eta_{10}|^2+K_2|\eta_{20}|^2+ \gamma_1(|\eta_{10}|^2-|\eta_{20}|^2) \right)
(H-H_{c2}+\delta B) \right.\nonumber\\
&\left.    + 2\beta_1(|\eta_{10}|^4+|\eta_{20}|^4)+2\beta_2|\eta_{10}|^2|\eta_{20}|^2     \right \rangle=0.
\label{eq}
\end{eqnarray}
For the more compact presentation I introduce the following notations for the coordinate dependent combinations
\begin{eqnarray}
&C_2({\bf r})=\frac{2\pi}{\Phi_0}\left(K_1|\eta_{10}|^2+K_2|\eta_{20}|^2\right )+
\gamma_1(|\eta_{10}|^2-|\eta_{20}|^2),\\
&C_4({\bf r})= 2\beta_1(|\eta_{10}|^4+|\eta_{20}|^4)+2\beta_2|\eta_{10}|^2|\eta_{20}|^2 
\end{eqnarray}
and rewrite (\ref{eq}) as
\begin{eqnarray}
\langle
C_2({\bf r})\rangle
(H-H_{c2})+\langle
C_2({\bf r})\delta B({\bf r})\rangle+\langle C_4({\bf r})\rangle
=0.
\label{eq1}
\end{eqnarray}
Hence, below the upper critical field 
 the magnetization decrease is
\begin{equation}
{\cal M}-M_0=\frac{\langle\delta B({\bf r})\rangle}{4\pi}=-\langle C_2({\bf r})-m({\bf r})
\rangle=-\frac{\langle C_2({\bf r})\rangle\langle C_2({\bf r})-m({\bf r})
\rangle}{\langle C_2({\bf r})\delta B({\bf r})\rangle+\langle C_4({\bf r})\rangle}(H_{c2}-H).
\label{A1}
\end{equation}
The pre-factor $\frac{\langle C_2({\bf r})\rangle\langle C_2({\bf r})-m({\bf r})
\rangle}{\langle C_2({\bf r})\delta B({\bf r})\rangle+\langle C_4({\bf r})\rangle}$ in the right hand side of this equation  plays the role of the generalized Abrikosov combination
\begin{equation}
\frac{1}{4\pi\beta_A(2\kappa^2-1)},
\label{A_2}
\end{equation}
where $\kappa$ is the Ginzburg-Landau parameter, and  $\beta_A$ is the Abrikosov constant. In one band superconductor, 
where the type (\ref{L1}) solution of the linear GL equation is $\eta_0({\bf r})$, this constant
$$
\beta_A =\frac{\langle|\eta_0({\bf r})|^4\rangle}{(\langle|\eta_0({\bf r})|^2\rangle)^2}
$$
is just the number independent from the material properties.
In two band case the universality is lost.
Taking in mind the Eqs.(\ref{deltaB}) and (\ref{m2})
\begin{equation}
\delta B({\bf r})=-4\pi \left [C_2({\bf r})-m({\bf r})\right],
\end{equation}
\begin{equation}
m({\bf r})=-\frac{2\pi C_2({\bf r})
}{\tilde\alpha +6\beta M_0^2}
\end{equation}
 we see that the pre-factor in Eq.(\ref{A1})  is expressed
 through the averages of $|\eta_{10}({\bf r})|^2$, $|\eta_{20}({\bf r})|^2$ and squares of them. The explicit calculation of it  can be performed only 
 after determination of constant $C_1$ and $C_2$ as stationary values of the GL functional taken at functions Eq.(\ref{L1}).
  
 The magnetic moment decrease in the ferromagnetic superconducting mixed state  is registered experimentally in URhGe \cite{Aoki01} and in UCoGe
 \cite{Paulsen2012,Hykel2014}. The temperature dependence of magnetization in URhGe is  shown in Fig.3.

\section{Transition from  superconducting to superconducting ferromagnetic state} 

In previous chapter we  discussed the phase transition from the ferromagnetic state to ferromagnetic superconducting mixed state
taking place in all uranium ferromagnets at temperature decrease in zero field (Fig.1) and also in  an external field parallel to the spontaneous magnetization. This case the superconducting order parameter forms the vortex lattice where  vortices are closely packed together: the distance between them is of the order of the coherence length $\xi(T)$.
 Another situation is realized in UCoGe (Fig.1c).  At pressures larger 1 GPa  the Curie temperature falls below the superconducting critical temperature and
 the phase transition occurs  to nonmagnetic superconducting state. The pressure decrease transforms this state into ferromagnetic superconducting state.
  Theoretically the phase transformation from normal to superconducting state and the subsequent transition to superconducting ferromagnet state in neutral superfluid with triplet pairing have been described in Ref.8,9.
 There was also predicted  \cite{Raghu2016}
  the direct first-order phase transition between the normal and superconducting ferromagnet state. It occurs in some pressures interval when the temperatures of transition to ferromagnetic and superconducting state are closed each other  (see Appendix).  So long the magnetization 
  is small enough  it
  does not penetrate inside the bulk of material being screened by the surface supercurrents.
  At pressure decrease the magnetization free superconducting state  passes 
  into the ferromagnetic superconducting mixed state. 
   This transformation is  complete analog  of transition between the Meissner and the mixed superconducting state \cite{Mineev2017}
   (see Fig.4).
    The ferromagnetic magnetization increasing with pressure decrease  penetrates to the superconducting volume in form of quantized vortices. This is happen when it reaches
 the value of the lower critical field $M_{c1}=H_{c1}/4\pi$ in this material.
 In the type-II ferromagnetic superconductors $H_{c1}\ll H_{c2}$ and at $M$ slightly above $M_{c1}$ the distance between vortices  
\begin{equation}
r_0\approx \xi\sqrt{\frac{H_{c2}}{4\pi M_{c1}}}
\end{equation}
is large in comparison with coherence length. Thus, it is reasonable to study 
  the field and the order parameter distributions around an isolated vortex.

\subsection{Single vortex}

An isolated vortex in an uniaxial metal  that I discuss is axially symmetric. It has a phase which changes by $2\pi$ after rotation around its axis directed along the spontaneous magnetization $M_0\hat z$. When the coefficient  $\gamma_2=-|\gamma_2|$ is negative the phase difference between the superconducting order parameters is absent \cite{Mineev2017} and I put it equal to the azimuthal angle $\varphi$  in the cylindrical frame $(r,\varphi,z)$.
Thus, I will look for a solution of GL equations (\ref{GL1}), (\ref{GL2})  in the form
\begin{equation}
\eta_1=f_1( r )e^{i\varphi},~~~~~\eta_2=f_2( r )e^{i\varphi}.
\end{equation}
The vector potential has only a $\varphi$-component: ${\bf A}=(0,A_{\varphi},0)$, and the gauge invariant vector potential is
\begin{equation}
{\bf Q}=(0, A_{\varphi}-\frac{\hbar c}{2er},0).
\end{equation}
The GL equations are
\begin{eqnarray}
(\alpha_1+\gamma_1B)f_1-K_1\left [ \frac{\partial^2}{\partial r^2}+\frac{1}{r}\frac{\partial}{\partial r}- \frac{4e^2Q^2}{\hbar^2c^2}
 \right ]f_1+\gamma_2f_2+2\beta_1f_1^3
+\beta_2f_1f_2^2=0,
\label{GL3}
\\
(\alpha_1-\gamma_1B)f_2-K_2\left [ \frac{\partial^2}{\partial r^2}+\frac{1}{r}\frac{\partial}{\partial r}- \frac{4e^2Q^2}{\hbar^2c^2}
 \right ]f_2+\gamma_2f_1+2\beta_1f_2^3
+\beta_2f_2f_1^2=0
\label{GL4}
\end{eqnarray}

The field distribution around a single vortex is determined by the Maxwell equation derived from the stationary condition 
of the GL functional with respect of vector potential
\begin{equation}
curl~curl {\bf A}-4\pi curl{\bf M}+4\pi \gamma_1 curl\hat z(f_1^2-f_2^2)+\frac{2(4\pi e)^2}{\pi(\hbar c)^2}(K_1f_1^2+K_2f_2^2){\bf Q}=0
\label{Ind}
\end{equation}
For $r\ne 0$ it is
\begin{equation}
curl~curl {\bf Q}-4\pi curl{\bf M}+4\pi \gamma_1curl\hat z(f_1^2-f_2^2)+\frac{2(4\pi e)^2}{\pi(\hbar c)^2}(K_1f_1^2+K_2f_2^2){\bf Q}=0
\label{Q}
\end{equation}
or
\begin{equation}
\frac{\partial}{\partial r}\frac{1}{r}\frac{\partial rQ}{\partial r} -\frac{2(4\pi e)^2}{\pi(\hbar c)^2}(K_1f_1^2+K_2f_2^2){Q}=
4\pi \frac{\partial( M-\gamma_1(f_1^2-f_2^2))}{\partial r}
\label{Lond}
\end{equation}
The magnetization  is determined from the equation
\begin{equation}
2\alpha M+4\beta M^3  -2D\Delta M-B
=0,
\end{equation}
obtained by the variation of the  functional Eq.(\ref{FE}) in  respect to $M$. 
The induction is 
$B=curl_z{\bf A}$.
Thus, omitting
 the  gradient term  $D\Delta M$   which can be thrown out by the same reason as in Eq.(\ref{m2}), we come to the  equation
\begin{equation}
2\alpha M+4\beta M^3
 =B=\frac{1}{r}\frac{\partial rQ}{\partial r},
\label{MSC}
\end{equation}
which is valid at $r\ne 0$.

The equations (\ref{GL3}), (\ref{GL4}), (\ref{Lond}) and (\ref{MSC})    present the full system of equations determining the space distribution of the $f_1( r ),~f_2( r ),
M( r )$, and $Q( r )$ around single vortex in the ferromagnetic superconductor. The solution of this system can be found only numerically. However, a qualitative description is still possible.

The general solution of  Eq.(\ref{Lond})
\begin{equation}
Q( r )=Q_h( r )+Q_i( r )
\end{equation}
consists of the sum of a solution of homogeneous equation and a particular solution of the inhomogeneous equation.

At distances  larger than the London penetration depth from the vortex axis$$r>\lambda =\frac{\sqrt{\pi}\hbar c}{4\pi e\sqrt{2(K_1f_{10}^2+K_2f_{20}^2)}}$$ 
the functions $f_1( r )\approx f_{10},~f_2
( r )\approx f_{20} , M( r )\approx M_0$ 
are almost constant and  
\begin{equation}
Q_h( r )=-\frac{\Phi_0}{2\pi\lambda}{\cal K}_1\left(\frac{r}{\lambda}\right),
\end{equation}
where the function ${\cal K}_1(z)$ is the Macdonald function of first order.
It decreases exponentially for large $z$:
$$
{\cal K}_1(z)=\sqrt{\frac{2}{\pi z }}\exp(-z).
$$
The constant magnetization $M_0$ is determined from the Eq.(63) with $B=0$.
The corresponding solution of inhomogeneous Eq.(\ref{Lond}) is
\begin{equation}
Q_i=4\pi \lambda^2\frac{\partial( M_0-\gamma_1(f_{10}^2-f_{20}^2))}{\partial r}=0.
\end{equation}
The induction $B=curl_z({\bf Q}_h+{\bf Q}_i)$ is exponentially small. The constants $ f_{10},~f_{20}$ are found from the equations (\ref{GL3}), (\ref{GL4})  
at $Q( r )=0$.

The solution of equations (\ref{GL3}), (\ref{GL4}) at the small $ r< \xi\approx \sqrt{K_{1,2}/|\alpha_1|}$
 is $f_1\propto r/\xi,~~~ f_2\propto r/\xi$. The induction $B=curl_z({\bf Q}_h+{\bf Q}_i)=B_0$, where the constant $B_0$ must be found as the limiting value
of the numerical solution of equations in intermediate region $\xi<r<\lambda$, $Q=2\pi rM$
and magnetization is determined by equation
\begin{equation}
2\alpha M+4\beta M^3
=B_0.
\end{equation}

 The crucial difference with vortex solution for ordinary type-II superconductors is the behavior of the order parameters in the intermediate
 distance interval $\xi<r<\lambda$. Here, 
  all the functions $f_1( r ),~f_2( r ),~M( r ),~ B (r )=curl_z{\bf A}$ are gradually changed (see Fig.5).
  
  \subsection{Lower critical field}
  
 The free energy of single vortex is the difference between the energy  Eq.(\ref {FE}) at stationary   vortex solution and the energy  without vortex, that is at stationary constant $\eta_1,~\eta_2, M_0, B=0$,
 \begin{eqnarray}
&E_v=\int dV\left\{\alpha M^2({\bf r})+\beta M^4({\bf r}) 
+\alpha_1(|\eta_1({\bf r})|^2+|\eta_2({\bf r})|^2)+\gamma_1B({\bf r})(|\eta_1({\bf r})|^2
-|\eta_2({\bf r})|^2)\right.\nonumber\\
&+\left.\gamma_2(\eta_1({\bf r})\eta_2^\star({\bf r})+\eta_1^\star({\bf r})\eta_2({\bf r}))+
\beta_1(|\eta_1({\bf r})|^4+|\eta_2({\bf r})|^4)+
\beta_2|\eta_1({\bf r})|^2|\eta_2({\bf r})|^2\right.\nonumber\\
&\left.+K_{1}[(D_x\eta_1({\bf r}))^\star D_x\eta_1({\bf r})+(D_y\eta_1({\bf r}))^\star D_y\eta_1({\bf r})]+K_{2}[(D_x\eta_2({\bf r}))^\star D_x\eta_2({\bf r})+(D_y\eta_2({\bf r}))^\star D_y\eta_2({\bf r})]+
\frac{{\bf B}^2}{8\pi}-{\bf B}{\bf M}\right \} \nonumber \\
&-\int dV\left\{\alpha M_0^2+\beta M_0^4 
+\alpha_1(|\eta_1|^2+|\eta_2|^2)
+\gamma_2(\eta_1\eta_2^\star+\eta_1^\star\eta_2)+
\beta_1(|\eta_1|^4+|\eta_2|^4)+
\beta_2|\eta_1|^2|\eta_2|^2\right\}
\label{VE}
\end{eqnarray} 
The corresponding expression for the conventional single band type-II superconductor  is obtained if we put $M=0,\\\eta_1=\eta_2$.
 This case the kinetic energy term contains $K_1 (4e^2/c^2)Q^2|\eta_1|^2$. Since $Q\propto1/r$ for $\xi\ll r\ll\lambda$, this gives a logarithmically large contribution at distances $r\sim\lambda$. Because modulus of the  vortex order parameter $|\eta_1({\bf r})|=|\eta_{1}|=const$
  everywhere at $r>\xi$ from the vortex axis the other terms  add nothing to the vortex energy. As result  the energy of a single-quantum Abrikosov vortex is
$$
  E_{vA}=\frac{\Phi_0^2}{(4\pi\lambda)^2}\ln\left (\frac{\lambda}{\xi}   \right ).
$$
 In ferromagnetic two-band superconductor with triplet pairing the situation is different. 
 In the interval of distances $\xi< r<\lambda$ all the order parameters $f_1( r ),~f_2( r ),~M( r )$ do not coincide with its values in the vortex absence.
 Hence, the vortex energy does not have the usual logarithmic form. It can be calculated only numerically making use the solution of Eqs.(\ref{GL3}), (\ref{GL4}), (\ref{Lond}) and (\ref{MSC}). 
 
 The free energy of a unit volume of a superconductor with set of single-quantum vortices is obtained by multiplication of the vortex energy on the
 density of vortices $n=\langle B\rangle/\Phi_0$, where $\langle B\rangle$ is the induction space average. 
 Magnetization begins penetrate in the bulk of superconductor when  loss of energy due to vortices appearance  will be compensated by gain of the energy 
 due to disappearance of work on pushing out of magnetization from  volume of superconductor 
 \begin{equation}
 \frac{\langle B\rangle}{\Phi_0}E_v-M\langle B\rangle<0.
 \end{equation}
Thus, in UCoGe,  at pressure decrease the magnetization reaches the lower critical value 
\begin{equation}
M_{c1}(P,T)=\frac{E_v}{\Phi_0},
 \end{equation}
and the transition from the Meissner to the superconducting mixed state occurs.  
In the presence of external field  parallel  to the domain magnetization this formula acquires the following form
\begin{equation}
(H+4\pi M)_{c1}=\frac{4\pi}{\Phi_0}E_v.
 \end{equation}

\section{Conclusion}
  
 I have developed the theory of type-II superconductivity in two band ferromagnetic metals with triplet pairing. 
 The obtained results near the upper critical field are   in qualitative correspondence with the results of classic  Abrikosov theory for type-II superconductivity in  single band  metals with singlet pairing. However, the  magnetization decrease below the transition to the superconducting ferromagnetic state is not expressed through the universal ratio known in the Abrikosov theory.
 The essential distinction also presents the coordinate dependence of the order parameters and the magnetic field around isolated quantized vortex that leads to the
 different magnitude in vortex line energy in comparison with its value in
 conventional superconductors.

The theory is applicable to the description  of superconducting state arising deeply inside the ferromagnetic state in UGe$_2$, URhGe, UCoGe.
The particular attention is devoted to the transition from the Meissner to the superconducting mixed state specific for UCoGe.

The presented approach can be also applied to the description of type-II superconductivity in two band nonmagnetic metals either with singlet or with triplet pairing.

\appendix
\section{}

The  direct first order transition from normal to superconducting ferromagnetic state in neutral Fermi liquid has been predicted by Cheung and Raghu \cite{Raghu2016} by means the numerical calculations. An attempt to confirm this  by analytical treatment undertaken in Ref.9 is incorrect. 
The proper qualitative argumentation in support of conclusion  Ref.8 is as follows. Taking electron charge equal to zero $e=0$ or, in other words, the London penetration depth equal to infinity  we come from the present model to the 
neutral Fermi liquid model discussed in Ref.8,9. This case according to the Eq.(\ref{Ind}) the magnetic induction is
\begin{equation}
B=4\pi M-4\pi\gamma_1(f_1^2-f_2^2).
\end{equation}
In absence of gradient terms the free energy density of the ferromagnetic superconductor in respect to the free energy density in the normal state is  
\begin{equation}
F=\alpha_0(T-T_c) M^2+\beta M^4 
+\alpha_1(\eta_1^2+\eta_2^2)+4\pi \gamma_1M(\eta_1^2-\eta_2^2)-
2|\gamma_2|\eta_1\eta_2+
\beta_1(\eta_1^4+\eta_2^4)+
\beta_2\eta_1^2\eta_2^2-2\pi[\gamma_1(\eta_1^2-\eta_2^2]^2 .
\end{equation}
In the normal state $\eta_1=\eta_2=M=0$ and $F=0$.  However, due to the linear in $M$
 term $4\pi \gamma_1M(\eta_1^2-\eta_2^2)$
one can find that the state with $F=0$ can be realized also at nonzero order parameter values  $\eta_1\ne 0,~\eta_2\ne 0,~M\ne 0$. 
These two states are divided by the phase transition of the first order.
 Indeed,  as this was shown in Ref.8,  the first order type transition occurs  near the intersection  the line $\alpha(T,P)=0$ with the line $ \alpha_1(T,P)=0$ . The width of pressures interval where the first order transition occurs is in fact negligibly small. This is due to the  smallness of $\gamma_1$ coefficient
 already pointed out  in the main text  (see Eq.(\ref{gamma1})). Here,
\begin{equation}
\frac{\gamma_1M}{\alpha_{10}}\approx\frac{\mu_BM}{\varepsilon_F}T_{sc0}\ll T_{sc0},
\end{equation}
where $\mu_B$ is the Bohr magneton and $\varepsilon_F$ is the Fermi energy \cite{Book}. 
Thus, the corresponding term is practically insignificant.

In charged Fermi liquid the  direct   transition from the normal to the superconducting ferromagnetic state will be apparently of the second order because the appearance of a finite magnetization accompanied by  work on pushing  out of the magnetic induction
from the superconducting volume.

\begin{figure}[p]
\includegraphics
[height=.8\textheight]
{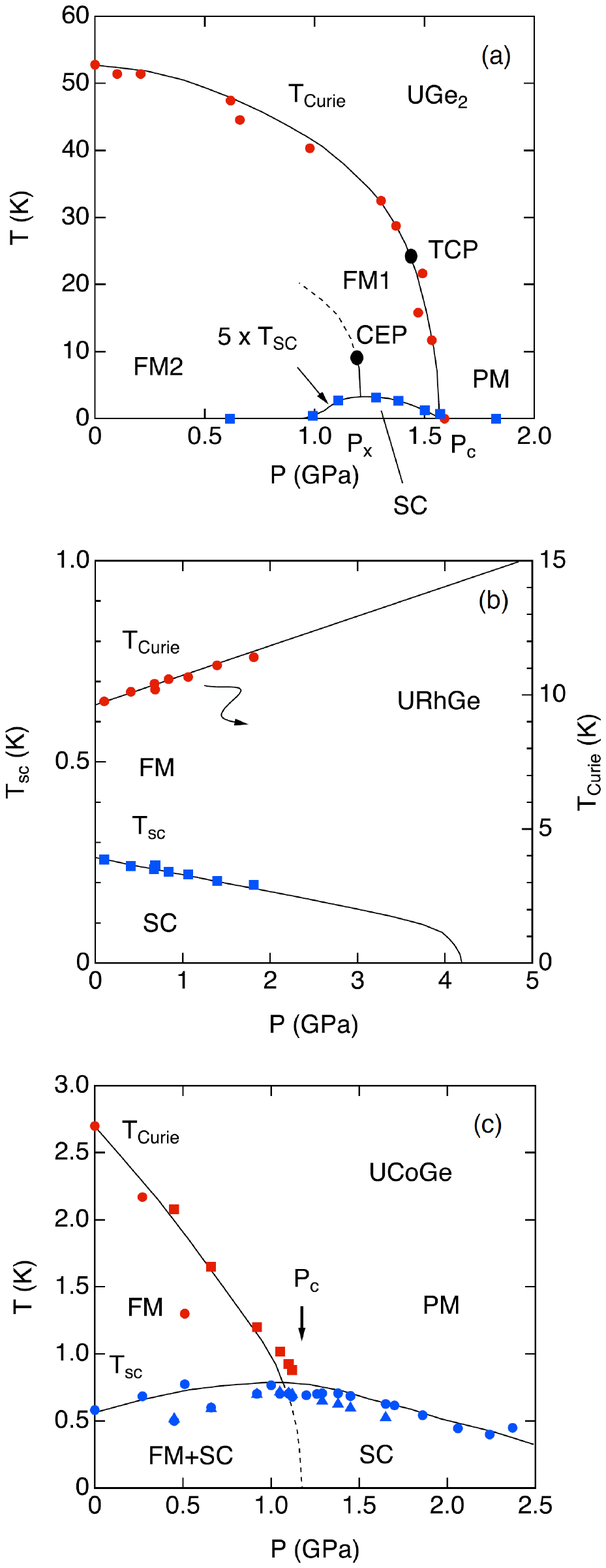}
 \caption{(Color online) Temperature-pressure phase diagram of UGe2,
URhGe, and UCoGe. Notations  FM, SC and PM have been used for ferromagnetic, superconducting and paramagnetic  phases correspondingly, TCP is the tricritical point, CEP is the critical end point.
\cite{Aoki2014} }
\end{figure}

\begin{figure}[p]
\includegraphics
[height=.8\textheight]
{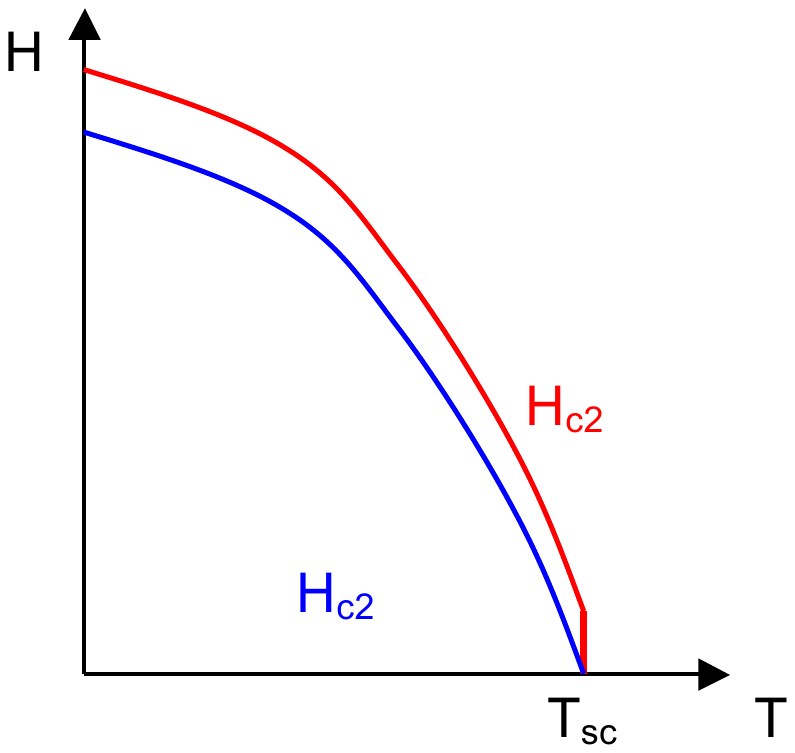}
 \caption{(Color online) Schematic upper critical field  $H_{c2}(T)$ temperature dependence: for single domain - lower (blue) curve
 and for multi-domain specimen - upper (red) curve.}
 \end{figure}

\begin{figure}[p]
\includegraphics
[height=.8\textheight]
{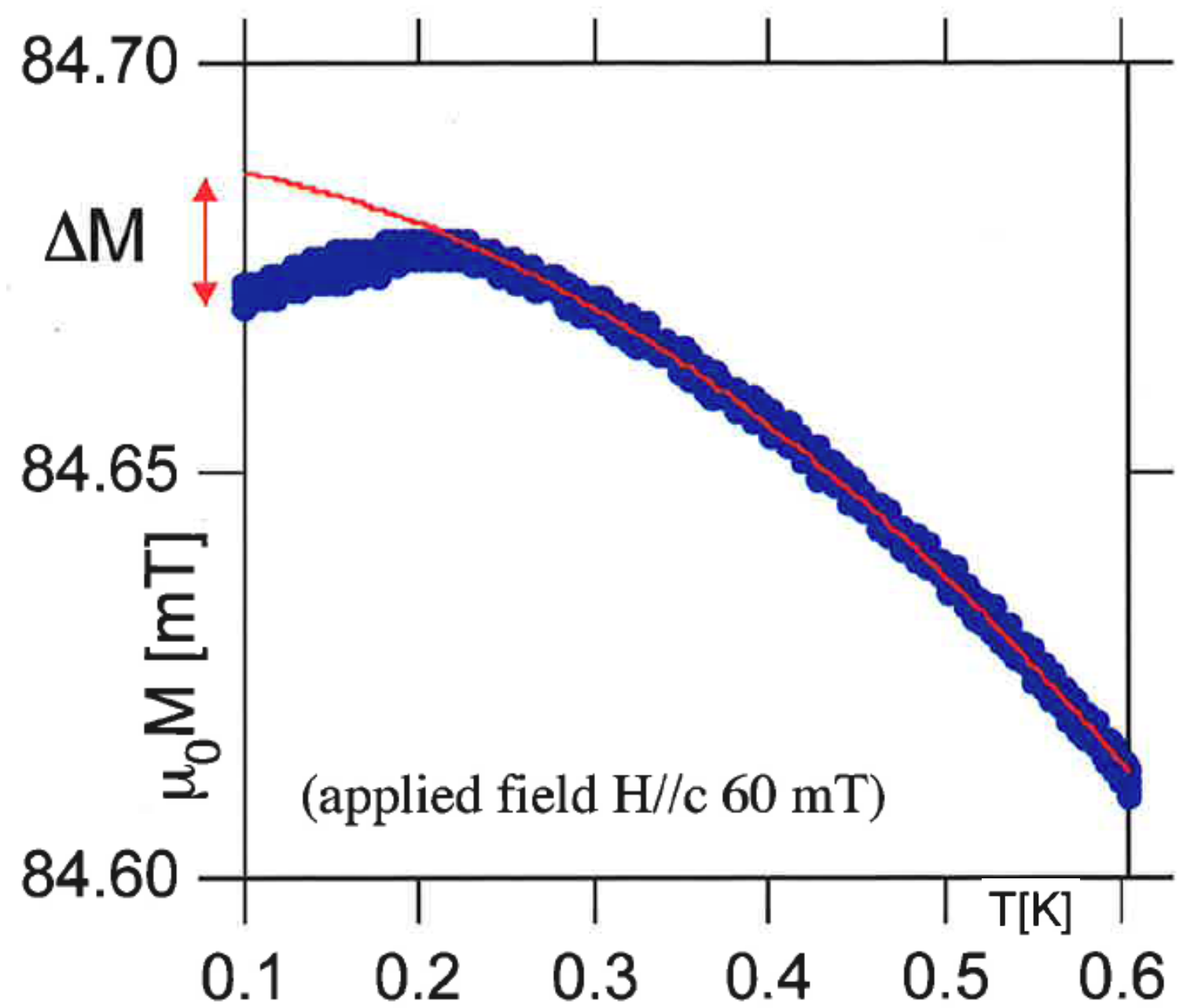}
 \caption{(Color online) The change of static magnetization in URhGe in a constant applied field of 0.06 T \cite{Huxley2004}.
 }
 \end{figure}

\begin{figure}[p]
\includegraphics
[height=.8\textheight]
{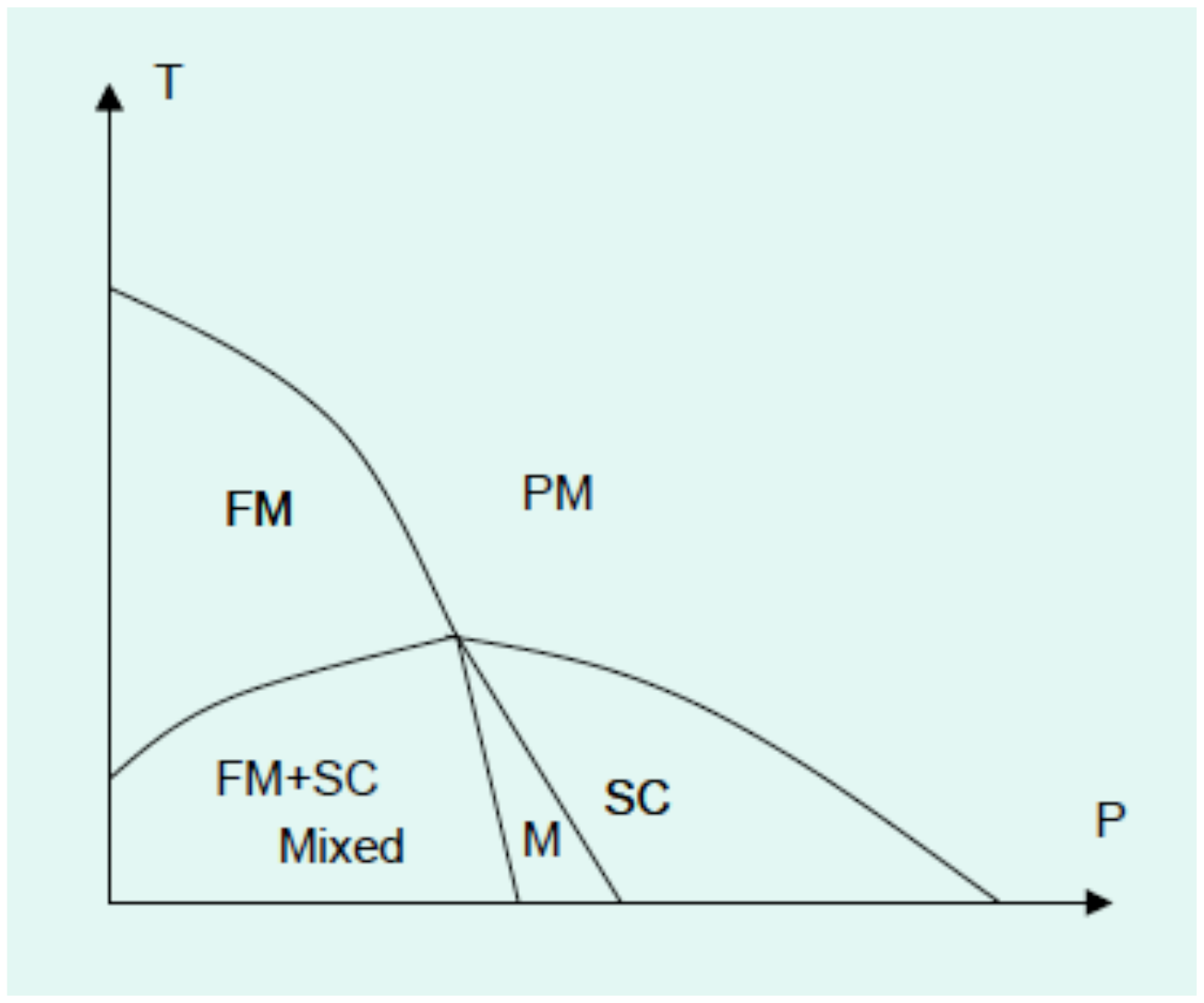}
 \caption{ Schematic temperature-pressure phase diagram of 
 UCoGe.
 Notations  FM, SC and PM  used for ferromagnetic, superconducting and paramagnetic  phases correspondingly. M is the Meissner state divided from the the mixed ferromagnetic superconducting states by the line of $H_{c1}$ 
  \cite{Mineev2017}.}
\end{figure}

 \begin{figure}[p]
\includegraphics
[height=.8\textheight]
{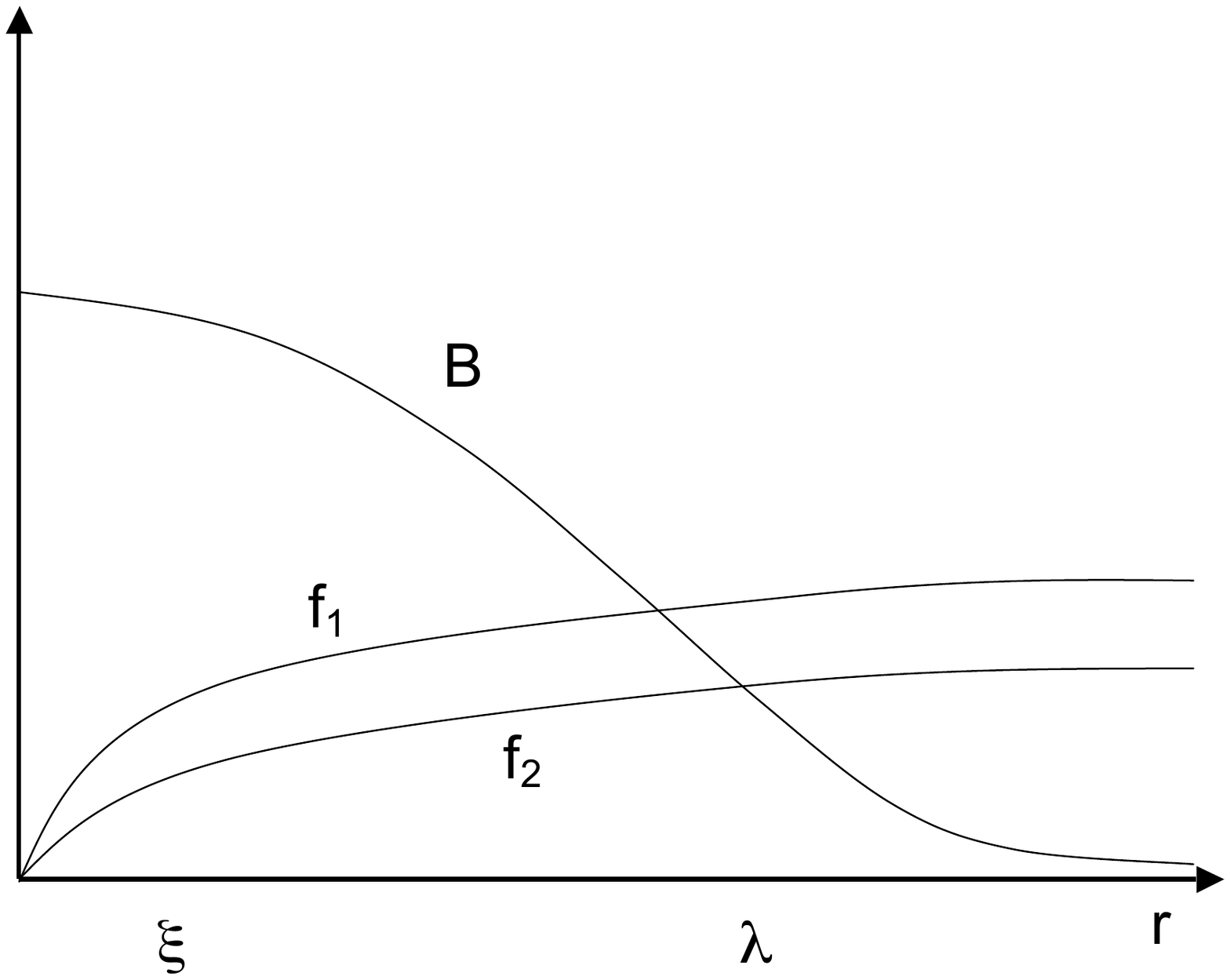}
 \caption{ Schematic  coordinate dependences around an isolated vortex of $f_1( r )$ and $f_2( r )$  superconducting order parameter amplitudes which
 grow up linearly at $r<\xi$ and tend to  constants at $r>\lambda$. $B( r )$ is the magnetic induction decreasing with distance from the vortex axis and
 tending to zero at $r>\lambda$. }
 \end{figure}

\end{document}